\documentclass[useAMS,usegraphicx, usenatbib]{mn2e}
\usepackage{times}
\usepackage{aas_macros}

\usepackage{amsmath}
\usepackage{amssymb}
\usepackage{mathtools}
\usepackage[utf8x]{inputenc}
\usepackage{graphicx}
\usepackage{epsfig}
\usepackage{latexsym}
\usepackage{color}
\usepackage{dcolumn}
\usepackage{bm}
\usepackage{natbib}
\usepackage{setspace}
\usepackage{subfigure}
\usepackage{psfrag}

\definecolor{Red}{rgb}{1, 0, 0}
\definecolor{Green}{rgb}{0, 1, 0}
\definecolor{Blue}{rgb}{0, 0, 1}
\definecolor{Black}{rgb}{0, 0, 0}
\definecolor{Grey}{rgb}{0.5, 0.5, 0.5}
\definecolor{White}{rgb}{1, 1, 1}
\definecolor{Yellow}{rgb}{1, 1, 0}
\definecolor{Magenta}{rgb}{1, 0, 1}
\definecolor{Cyan}{rgb}{0, 1, 1}
\definecolor{myCyan}{rgb}{0, 0.6, 1}
\definecolor{Orange}{rgb}{1, 0.5, 0}
\definecolor{Violet}{rgb}{0.5, 0, 0.5}
\definecolor{DarkRed}{rgb}{0.5, 0., 0.3}
\definecolor{Pink}{rgb}{1., 0.4, 0.7}
\definecolor{LightPink}{rgb}{1, 0.8, 0.8}
\definecolor{YellowGreen}{rgb}{0.6, 0.8, 0}
\definecolor{LightYellow}{rgb}{1., 1., 0.95}
\definecolor{Brown}{cmyk}{0, 0.8, 1, 0.6}

\begin{document}
\pagestyle{myheadings}
\markboth{Analytical photo-\emph{z} determination}{}

\title{The first analytical expression to estimate photometric redshifts suggested by a machine}
\author[A.~ Krone-Martins, E.E.O. Ishida, R.~S.~de Souza]
{A.~ Krone-Martins$^{1}$, E.E.O. Ishida$^{2,3}$, R. S. de Souza$^{4,5}$
\\
$^{1}$SIM, Faculdade de Ci\^encias, Universidade de Lisboa, Ed. C8, Campo Grande, 1749-016, Lisboa, Portugal\\
$^{2}$IAG, Universidade de S\~{a}o Paulo, Rua do Mat\~{a}o 1226,  05508-900, S\~{a}o Paulo, SP, Brazil\\
$^{3}$Max-Planck-Institut f\"ur Astrophysik, Karl-Schwarzschild-Str. 1, D-85748 Garching, Germany\\
$^{4}$Korea Astronomy \& Space Science Institute, Daedeokdae-ro 776,  305-348  Daejeon, Korea\\
$^{5}$MTA E\"otv\"os University, EIRSA "Lendulet" Astrophysics Research Group, Budapest 1117, Hungary
}


\pagerange{0000--00000} \pubyear{2013}

\maketitle
\label{firstpage}


\topmargin -1.3cm


\maketitle

\begin{abstract}
We report the first analytical expression purely constructed by a machine  to determine photometric redshifts ($z_{\rm phot}$) of galaxies. A simple and reliable functional form is derived  using $41,214$ galaxies from the Sloan Digital Sky Survey Data Release 10 (SDSS-DR10) spectroscopic sample. The method automatically dropped the $u$ and $z$ bands, relying only on $g$, $r$ and $i$ for the final solution. Applying this expression to other $1,417,181$ SDSS-DR10 galaxies,  with measured spectroscopic redshifts ($z_{\rm spec}$), we achieved a mean $\langle (z_{\rm phot} - z_{\rm spec})/(1+z_{\rm spec})\rangle\lesssim 0.0086$ and a scatter $\sigma_{(z_{\rm phot} - z_{\rm spec})/(1+z_{\rm spec})}\lesssim 0.045$ when averaged up to $z \lesssim 1.0$. The method was also applied to the PHAT0 dataset, confirming  the competitiveness of our results when faced with other methods from the literature. This is the first use of symbolic regression in cosmology, representing a leap forward in astronomy-data-mining connection.
\end{abstract}


\section{Introduction}
\label{sec:intro}

A novel methodology was recently proposed to automatically search for underlying analytical laws in data \citep{schmidt2009}. Its importance has been highlighted into astronomy by \citet{graham2013}, and this letter is the first attempt to use it in a cosmological context.  We applied the aforementioned method to derive an analytic expression for photometric redshift (photo-\emph{z}) determination from Sloan Digital Sky Survey 10th data release  spectroscopic sample of galaxies \citep[SDSS-DR10,][]{ahn2013}. Our goal here is to  demonstrate the potential of machine proposed analytical relations in providing  simple and reliable photo-\emph{z}.

Due to the variety of spectra occurring in nature (as there are several types of galaxies, of different ages, metallicities, star-forming histories, merging histories, etc.), the unicity of photometric redshift estimates is not assured for any sample. Nevertheless, the large amount of data expected to be observed by surveys like the Large Synoptic Survey Telescope\footnote{http://www.lsst.org/lsst/} \citep{lsst},  EUCLID\footnote{http://sci.esa.int/euclid/} \citep{euclid} or  Wide-Field Survey Infrared Telescope\footnote{http://wfirst.gsfc.nasa.gov/} \citep{wfirst} makes it infeasible to obtain spectroscopic redshifts for all their objects with the current and likely near future technology. 
Therefore making photo-\emph{z} is the only viable solution for estimating redshifts in such large scale.

Photo-\emph{z} methods have been widely used in fields as diverse as gravitational lensing \citep[e.g.,][]{Mandelbaum2008, zitrin2011,nusser2013}, baryon acoustic oscillations \citep[e.g.,][]{nishizawa2013}, quasars \citep[e.g.,][]{Richards2009}, luminous red galaxies  \citep[LRGs, ][]{desimoni2013} and supernovae \citep[e.g.,][]{Kessler2010}. At the same time, numerous efforts to accurately determine photo-\emph{z} were reported 
\citep[for a glimpse on the diversity of existent methods, see][and references therein]{abdalla2008, hildebrandt2010, zheng2012}. To deepen our understanding of the differences between photo-\emph{z} techniques, \citet{abdalla2008} compared results from six methods applied to  LRGs. They show 1-$\sigma$ scatters between 0.057 and 0.097 when averaged over the considered redshift range ($0.3 \leq z \leq 0.8$), systematically presenting poor accuracy at low ($z \leq$ 0.4) and high ($z\geq$ 0.7) redshifts. More recently, \citet{hildebrandt2010} presented a wider comparison enclosing 16 different methods. The methods perform better in simulated than real data, with empirical codes showing smaller biases than template-fitting ones.

The existing approaches are usually divided in two classes: empirical \citep[e.g.][]{connolly1995, collister2004, wadadekar2005, miles2007, omill2011, reis2012, carrasco2013} and template-fitting-based methods \citep[e.g.,][]{benitez2000,bolzonella2000,ilbert2006}. The former uses magnitudes and/or colours of a spectroscopically measured sample for training the method, which is then applied to the photometric sample. The latter, try to find spectral template and redshift which best fit the photometric observations using a library of well know observational or synthetic spectra.

The main advantage of the approach adopted in this letter is that without any \textit{a priori}  physical  information nor \textit{ad hoc} functional form, it empirically derives analytical expressions from the data. Besides that, the error propagation from the observables can be straightforwardly performed into the redshift estimate. Also, due to its analytic nature, the outcomes are more tractable, and thus interpretable, than the outcomes of other methods, such as neural-networks or support vector machines, for instance. Finally, the resulting expressions are promptly portable, and might be even incorporated on-the-fly via \textit{structured query language} (SQL) when retrieving catalogue data, for instance.

The outline of this letter is as follows. In Section \ref{sec:method}, we give a broad picture of the methodology followed in this letter. Then, section \ref{sec:data}, provides an overview of the adopted dataset. Afterwards, we present our results and compare with the recent literature in in  Section \ref{sec:res}. Finally, conclusions are presented in Section \ref{sec:conclusion}.

\section{Methodology}
\label{sec:method}

The ultimate goal of symbolic regression-based techniques is to find a functional form that explains hidden associations in datasets, while optimizing a given error metric \citep[e.g.,][]{schmidt2009}. This is fundamentally distinct from linear and non-linear regression methods that fit parameters for an \textit{a priori} analytical expression. In symbolic regression,  the machine searches the best expression and the optimal coefficients simultaneously. 

We used the software {\sc eureqa}\footnote{http://www.nutonian.com/products/eureqa/} \citep{schmidt2009} to test the application of symbolic regression for photo-\emph{z} determination. It allows the user to choose atomic function blocks (basic mathematical operations, exponentials, logarithms, boolean operators, trigonometric functions, etc.). Then, {\sc eureqa} scans through  the  data and a variety of combinations between the atomic function blocks are evolved through genetic programming \citep{Koza1992},  optimizing conciseness and accuracy. Lastly, the outcome functions are ordered  according to their complexity and quality of the fit.

The application of {\sc eureqa} to our problem follows a straightforward approach. First, a subset of galaxies with measured spectroscopic redshifts is used to derive an analytical expression that optimally predicts the redshift from the magnitude and colour data. In other words, an expression whose evaluation minimizes the mean absolute error when compared to the data. To seek simplicity while keeping accuracy, we only allowed the use of simple mathematical operations ({\tt +} , {\tt -} , {\tt *} , {\tt /}). Afterwards, the obtained expression is applied to a larger sample of galaxies with spectroscopic measurements, to perform a strict validation of the expression's predictive capability against real spectroscopic redshifts.


\section{Data}
\label{sec:data}

The data adopted in this work was selected from the SDSS-DR10 spectroscopic sample. This includes hundreds of thousands of new galaxies and quasars spectra from the Baryon Oscillation Spectroscopic Survey\footnote{http://www.sdss3.org/surveys/boss.php} (BOSS) in addition to all imaging and spectra from prior SDSS data releases. 

From this dataset, we selected all objects with spectroscopic measurements (table \texttt{SpecObj}) classified as galaxies (flag \texttt{SpecObj.class = 'GALAXY'}) and whose spectra were free from known problems (flag \texttt{SpecObj.zWarning = 0}). Moreover,  only sources with clean photometric measurements (flag \texttt{PhotoObj.CLEAN = 1}) were accepted. The SQL query used in SDSS CasJobs\footnote{http://skyserver.sdss3.org/casjobs/} service was

\begin{verbatim}
SELECT s.specObjID, g.u, g.g, g.r, g.i, g.z, 
       s.z AS redshift 
       INTO mydb.specObjAllz_cleanphoto 
FROM SpecObj AS s JOIN Galaxy AS g 
     ON s.specobjid = g.specobjid, PhotoObj
WHERE class = 'GALAXY' AND zWarning = 0 
      AND g.objId = PhotoObj.ObjID 
      AND PhotoObj.CLEAN=1
\end{verbatim}
where {\tt s.specObjID} is the object identification in the spectral tables, and {\tt g.u}, {\tt g.g}, {\tt g.r}, {\tt g.i}, {\tt g.z}, {\tt s.z} represent the SDSS' \textit{ugriz} model magnitudes and measured spectroscopic redshift, respectively. This resulted into a dataset containing $1,458,404$ objects, from which we retained only galaxies with $z_{\rm spec} < 1.0$. Additionally, all possible color combinations based on the available photometric bands were computed. 

We divided the data into two subsets, one for deriving the analytic expression and another for validation and error assessment. To mitigate biases created by unbalanced data, we randomly selected $5,000$ galaxies per redshift bin (width $\Delta z_{\rm spec} = 0.1$) up to $z_{\rm spec} = 0.8$. For $0.8 \leq z_{\rm spec} < 1.0$, half of all available objects in each redshift bin were used for deriving the expression. This comprises a total of $41,214$ galaxies that were used for searching the expression. Then, the accuracy (systematic errors) and precision (random errors) of this expression was assessed based on other $1,417,181$ objects. We only considered objects with $z_{\rm spec} > 0$.

Finally, we did not apply any cuts in magnitude, quality of spectroscopic redshift measurement nor galaxy types. This ensures that our results  are not biased towards high signal to noise data, a particular galaxy type nor optimal observation conditions in comparison with the SDSS-DR10 spectroscopic sample.

\section{Results}
\label{sec:res}

\begin{figure}
\includegraphics[width=82mm]{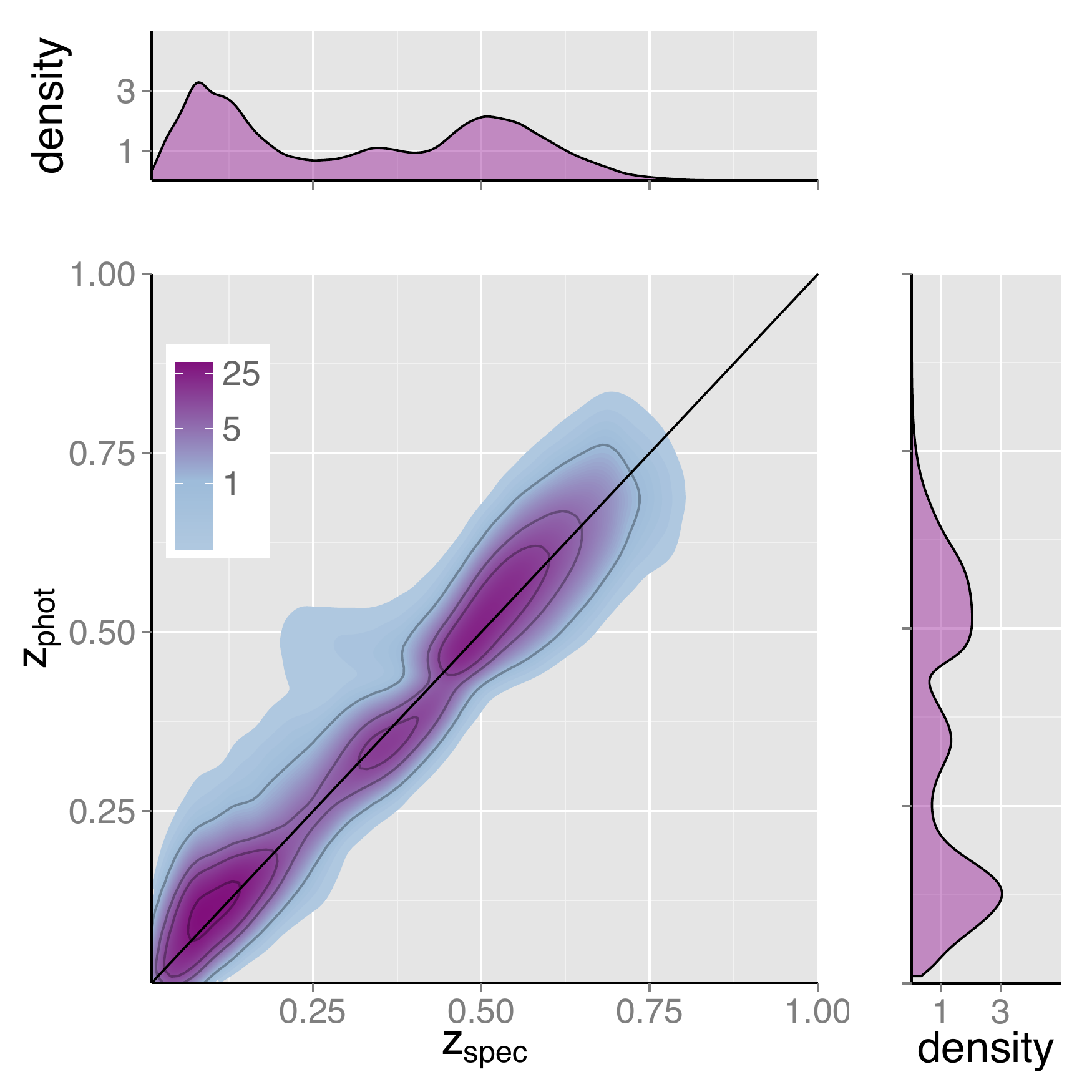}
\caption{Kernel density distribution of  photometric  ($z_{\rm phot}$) versus  spectroscopic  ($z_{\rm spec}$) redshifts for more than one million SDSS-DR10 galaxies. The colour  scale is logarithm, so a  difference of one is equivalent to a density variation by   a factor of $e$. Distributions for $z_{\rm spec}$ and $z_{\rm phot}$ redshifts are shown on the top and right panels.}
\label{figureRecMeasured}
\end{figure}

\begin{figure*}
\begin{minipage}[b]{0.975\linewidth}
\centering
\includegraphics[width=0.975\linewidth]{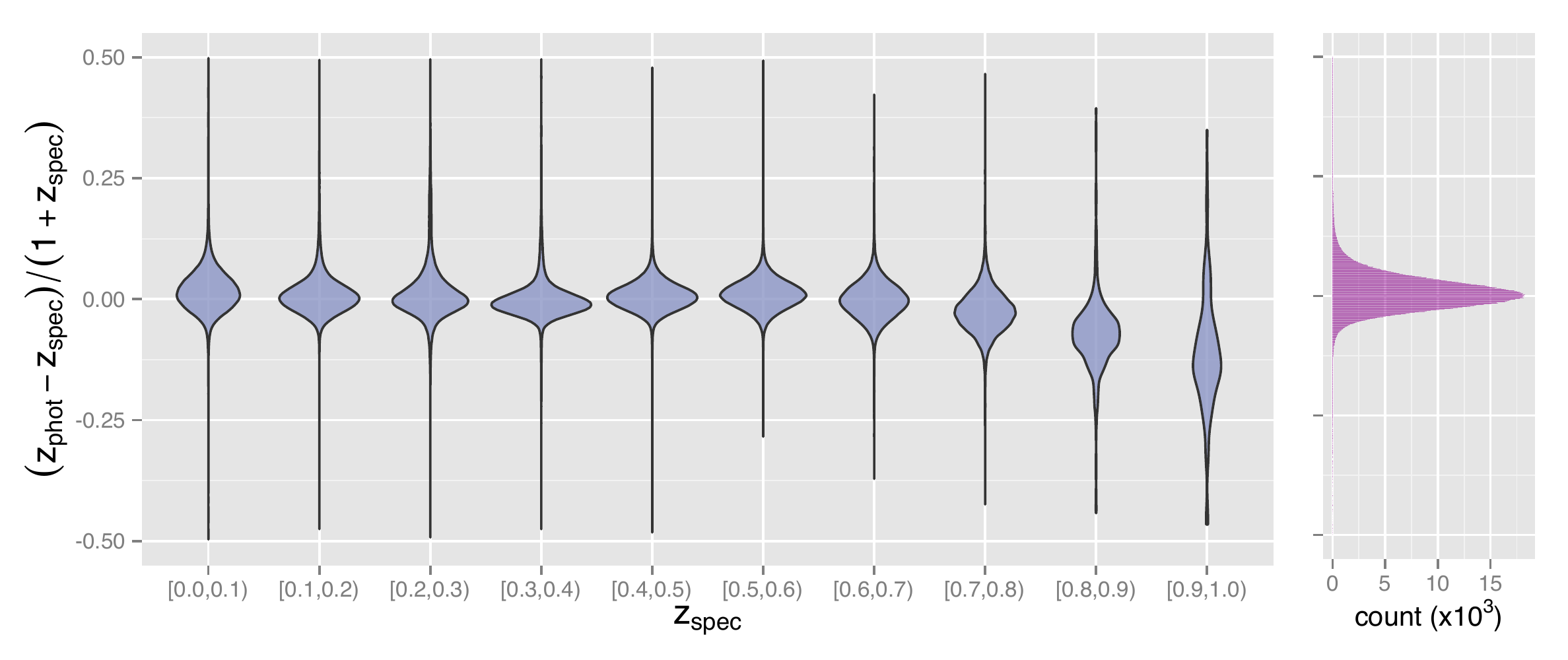}
\caption{Left panel shows the photometric redshift error distributions estimated from Eq. \ref{equationThePhotoZer}, in redshift bins of width $\Delta z_{\rm spec} = 0.1$. Right panel displays the error distribution for more than one million galaxies in SDSS-DR10 as a histogram with bins of width $\Delta((z_{\rm phot}-z_{\rm spec})/(1+z_{\rm spec})) = 0.001$.}
\label{figureErrDist}
\end{minipage}
\end{figure*}

Adding the ingredients described so far, the optimal functional form suggested by {\sc eureqa} to the adopted dataset is

\begin{equation}
\begin{split}
z_{\rm phot} = \frac{0.4436r - 8.261}{24.4 + (g-r)^2(g-i)^2(r-i)^2 - g}\\
+0.5152(r-i).
\end{split}
\label{equationThePhotoZer}
\end{equation}

This represents a rather simple empirical relation between photometric measurements and redshifts of galaxies calibrated for the SDSS-DR10 spectroscopic sample\footnote{We stress that this expression was calibrated for the SDSS-DR10 spectroscopic sample, and should not be extrapolated out of this scope.}. Given its analytical nature, Eq. \ref{equationThePhotoZer} allows a straightforward error propagation from the uncertainties in the measured magnitudes to the final photometric redshift.
Note the missing $u$ and $z$ bands  in the former equation. 
Such behavior was observed in several equations constructed by {\sc eureqa}, suggesting that a competitive performance might be reached using only three SDSS photometric bands\footnote{As a matter of comparison, \citet{hildebrandt2010} used 14 distinct bands, while \citet{abdalla2008} adopted all five SDSS bands.}.

Interestingly, the two SDSS filters kept out of the  derived equation are those which do not bracket the main spectral feature for imprinting redshift signature in photometry, for the redshift range considered in this work: the $\sim 4000$\AA ~break. This does not mean that these filters carry null information. Instead, it only highlights that the bulk of information relevant to photometric redshift determination relies on the other filters. Due to a compromise between error and complexity during the optimization procedure, only the most relevant filters survive to the output equations. Moreover, the expressions assembled by {\sc eureqa}  are not simply high order polynomials with additional terms, but more intricate combinations of magnitudes in different filters. Accordingly, expressions with more terms are not necessarily expected to improve redshift estimates, as additional terms might even introduce degeneracies.

To test the performance of Eq. (\ref{equationThePhotoZer}), we applied it to the photometric data of $1,417,181$ galaxies. Figure \ref{figureRecMeasured} summarizes our results, showing a comparison between $z_{\rm spec}$ and $z_{\rm phot}$. One can promptly notice that $z_{\rm spec}$ is well recovered by $z_{\rm phot}$ with reasonable accuracy. Furthermore, a reasonable match between $z_{\rm spec}$ and $z_{\rm phot}$ distributions can be observed (upper and right panels, respectively). This indicates that Eq. (\ref{equationThePhotoZer}) recovers the underlying redshift distribution over a significant fraction of the explored redshift range.

Left panel of figure \ref{figureErrDist} shows the probability  distribution functions  (PDF) of $(z_{\rm phot}-z_{\rm spec})/(1+z_{\rm spec})$ in each redshift bin (width $\Delta z_{\rm spec} = 0.1$) for $0\le z_{\rm spec} <1.0$, represented as violin plots. Each ``violin'' centre represents the median of the distribution, while the shape, its the mirrored PDF. The drop in medians at high redshifts ($z_{\rm spec} \gtrsim 0.7$) indicates that $z_{\rm phot}$ systematically underestimates $z_{\rm spec}$ at this range. This might be caused by poor statistics: in the full dataset, at $z_{\rm spec} \ge 0.8$ there are only $2,428$ objects, while for $z_{\rm spec} \ge 0.7$ there are $25,439$. This underweights the contribution of high-\emph{z} objects to the construction of Eq. (\ref{equationThePhotoZer}). Accordingly, for bins with equally balanced number of galaxies ($z_{\rm spec}\leqslant0.7$), no obvious systematic effects are seen.

Right panel of figure \ref{figureErrDist} shows a histogram of $(z_{\rm phot}-z_{\rm spec})/(1+z_{\rm spec})$,  with bins of $0.001$, forming a nearly perfect normal error distribution. As the mean and standard deviation are known to be sensitive to outliers, we removed the extreme tails of $z_{\rm phot}$ distribution prior to computing them (117 events, or less than 0.008 per cent of the sample). This rejection is performed directly in the $z_{\rm phot}$ distribution without any prior knowledge about $z_{\rm spec}$. The mean is $\langle (z_{\rm phot}-z_{\rm spec})/(1+z_{\rm spec}) \rangle \approx 0.0086$, while the scatter is $\sigma_{(z_{\rm phot}-z_{\rm spec})/(1+z_{\rm spec})} \approx 0.0449$.\footnote{A more robust statistical estimator against outliers are the median and median absolute deviation values (MAD). For this dataset, we obtained a median of $0.0048$ and $\textrm{MAD}=0.0318$.} Albeit using a different dataset, \citet{hildebrandt2010} obtained similar values ($0.005 \le \lvert \langle (z_{\rm phot}-z_{\rm spec})/(1+z_{\rm spec})\rangle \rvert \le 0.039$ and $0.034 \le \sigma_(z_{\rm phot}-z_{\rm spec})/(1+z_{\rm spec}) \le 0.076$). Nevertheless, given the different adopted datasets, we refrain from performing a direct comparison with our results. Notwithstanding, these figures suggest that equations derived by {\sc eureqa} might be competitive against more elaborated methods.

Using a homogeneous sample of LRGs,   \citet{abdalla2008} tested six different methods, reporting $0.0014 \le \lvert \langle z_{\textrm{phot}} - z_{\textrm{spec}}\rangle \rvert \le  0.0302$ and $0.0575 \le \sigma_{(z_{\textrm{phot}}-z_{\textrm{spec}})} \le 0.0973$. These values are compatible with those obtained by Eq. (\ref{equationThePhotoZer}), $\langle z_{\textrm{phot}} - z_{\textrm{spec}} \rangle \approx 0.0104$, with a scatter\footnote{Using robust statistics, we obtain a median of $0.0062$ and $\textrm{MAD}=0.0414$.} of $\sigma_{(z_{\textrm{phot}}-z_{\textrm{spec}})} \approx 0.0570$ . This reinforces the relevance of results achieved by the analytical expression derived with \textsc{eureqa}. Despite its simple nature, it was able to deliver competitive accuracy and precision from a rather diverse and inhomogeneous sample.

We have also explicitly searched for expressions 
incorporating the u or z filters. One example of such functional form is 

\begin{equation}
\begin{split}
z_{\rm phot} = 0.4583(r-i) + \frac{0.001i^2r - 0.3170r}{4.6691 + (u-i)(g-r)}.
\end{split}
\label{equationTheAddFilteredPhotoZer}
\end{equation}
Using this equation, we achieved accuracy and precision levels of $\langle (z_{\rm phot}-z_{\rm spec})/(1+z_{\rm spec}) \rangle \approx 0.0022$ and $\sigma_{(z_{\rm phot}-z_{\rm spec})/(1+z_{\rm spec})} \approx 0.0521$, respectively. These results are not better than those obtained with Eq \ref{equationThePhotoZer},  exemplifying  that a larger number of filters does not necessarily lead to a more accurate photometric redshift estimation.

\begin{figure}
\includegraphics[width=73mm]{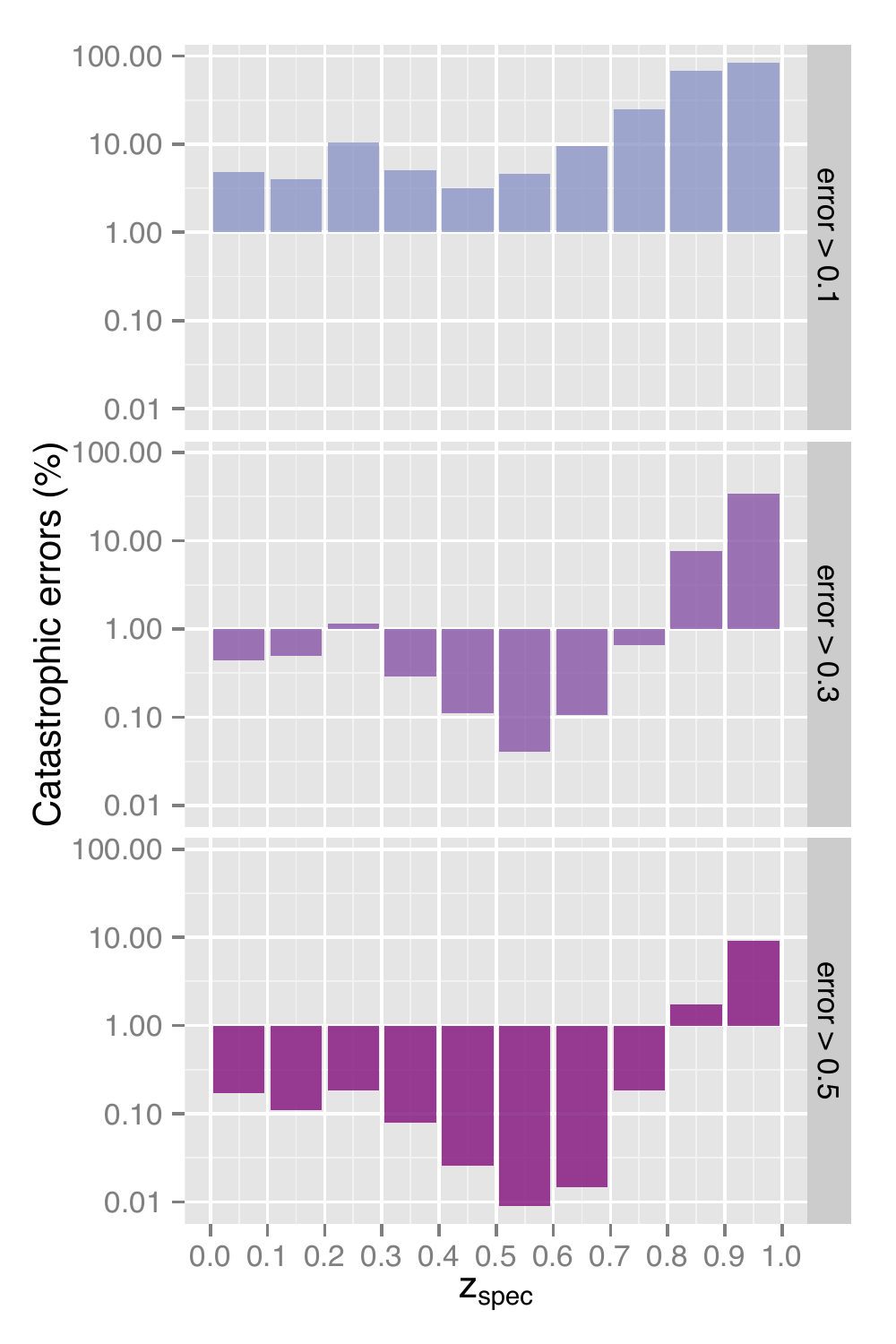}
\caption{Percentual of catastrophic errors resulting from the photo-\emph{z} estimation at each redshift bin for three different scenarios: $|z_{\rm phot} - z_{\rm spec}| > 0.1, 0.3$ and $0.5$, from top to bottom.}
\label{figureCatErr}
\end{figure}

To estimate the level of bias introduced by Eq. (\ref{equationThePhotoZer}) into a given cosmological inference, it is necessary to discuss the number of catastrophic errors, i.e., cases when photo-\emph{z} is above a given tolerance threshold \citep{bernstein2010}. These authors consider catastrophic errors as $\lvert z_{\rm phot} - z_{\rm spec} \rvert \gtrsim 1$, while \cite{hildebrandt2010} defined them as $\lvert z_{\rm phot} - z_{\rm spec} \rvert > 0.15 (1+z_{\rm spec})$ or $ > 0.5$. \citet{molino2013} consider redshift dependent limits in terms of median and MAD, which in our context means $\lvert z_{\rm phot} - z_{\rm spec} \rvert \geq 0.2$ at $z = 0$ and 0.39 at $z = 1.0$. Figure \ref{figureCatErr} shows the catastrophic error rate obtained from Eq. (\ref{equationThePhotoZer}) as a function of $z_{\rm spec}$ for three different scenarios: $\lvert z_{\rm phot} - z_{\rm spec} \rvert > 0.1$, $0.25$ and $0.5$. The choice of three independent criteria gives a glimpse of how Eq. (\ref{equationThePhotoZer}) performs in a wide range of accuracy requirements. In each panel the barplots are given in logarithm scale, where face-down bars indicate less than one percent of catastrophic errors according to the criteria on the right. 

\section{Conclusions}
\label{sec:conclusion}
 
This work is the first attempt to use a heuristic machine assistant to propose new analytical relationships for photo-\emph{z} estimation. It provides a simple and accurate functional form based on photometric information of SDSS spectroscopic sample galaxies. Although we started the search using all five SDSS bands, several solutions relied only on three of them. Hence, showing that for SDSS bands, a competitive performance can be attained even with a moderate number of filters.

We adopted a set of $41,214$ galaxies for determining the photo-\emph{z} expression. Afterwards, it was used to estimate $z_{\rm phot}$ for another $1,417,181$ galaxies with known $z_{\rm spec}$. Our results achieved  $\langle (z_{\rm phot} - z_{\rm spec})/(1+z_{\rm spec})\rangle\lesssim 0.0086$ and a scatter $\sigma_{(z_{\rm phot} - z_{\rm spec})/(1+z_{\rm spec})}\lesssim 0.045$ when averaged up to $z \lesssim 1.0$. These results indicate that symbolic regression is competitive against other methods available in the literature. An inspection of the $(z_{\rm phot}-z_{\rm spec})/(1+z_{\rm spec})$ distributions per redshift bin reveals systematic effects at $z_{\rm spec} \gtrsim 0.7$. Such behavior might be caused by the poor statistics at high redshifts.

The conciseness of the outcomes obtained by {\sc eureqa} are stressed by how easily they can be adopted by the astronomical community. The functions can even be directly incorporated into simple SQL queries. Such level of portability is unattainable by the majority of photo-\emph{z} methods currently available \citep[but see e.g.][]{connolly1995,hsieh2005}. Moreover, the error propagation can be straightforwardly achieved by deriving the redshift as a function of photometric observables \citep[e.g.,][]{collister2004,oyaizu2008}.

Finally, the possibility to use computers to unveil hidden analytical relationships in datasets, a heretofore task exclusive of humans, is astonishing \citep[e.g.][]{schmidt2009, graham2013}. Astronomy is already being flooded by an unprecedented amount of data, and this tendency is expected to increase even more in the next decade. Therefore the possibility to connect these novel systems to databases, and particularly allowing them to perform text mining in scientific literature \citep[as in][]{Leach2009}, might represent a new paradigm for astronomical exploration. These methods are coming to stay, and although still incipient and naive, they host a great potential to help humankind in its endeavour to unravel the Universe.


\section*{Acknowledgments}
We thank Reinaldo Ramos de Carvalho, Andressa Jendreieck, Laerte Sodr\'e Jr., Filipe Abdalla, Jon Loveday, Matias Carrasco, Jonatan D. Hernandez Fernandez and Ana Laura O'Mill for interesting suggestions and comments. EEOI and RSS thank the SIM Laboratory of the \textit{Universidade de Lisboa} for hospitality during the development of this work. This work was partially supported by the ESA VA4D project (AO 1-6740/11/F/MOS). AKM thanks the Portuguese agency \emph{Funda\c c\~ao para Ci\^encia e Tecnologia}, \emph{FCT}, for financial support (SFRH/BPD/74697/2010). EEOI thanks the Brazilian agencies FAPESP (2011/09525-3) and CAPES (9229-13-2) for financial support. Funding for SDSS-III has been provided by the Alfred P. Sloan Foundation, the Participating Institutions, the National Science Foundation, and the U.S. Department of Energy Office of Science. The SDSS-III web site is http://www.sdss3.org/.
This work was written on the collaborative ShareLaTeX platform.

\begin{figure*}
\begin{minipage}[b]{0.95\linewidth}
\centering
\includegraphics[width=0.95\linewidth]{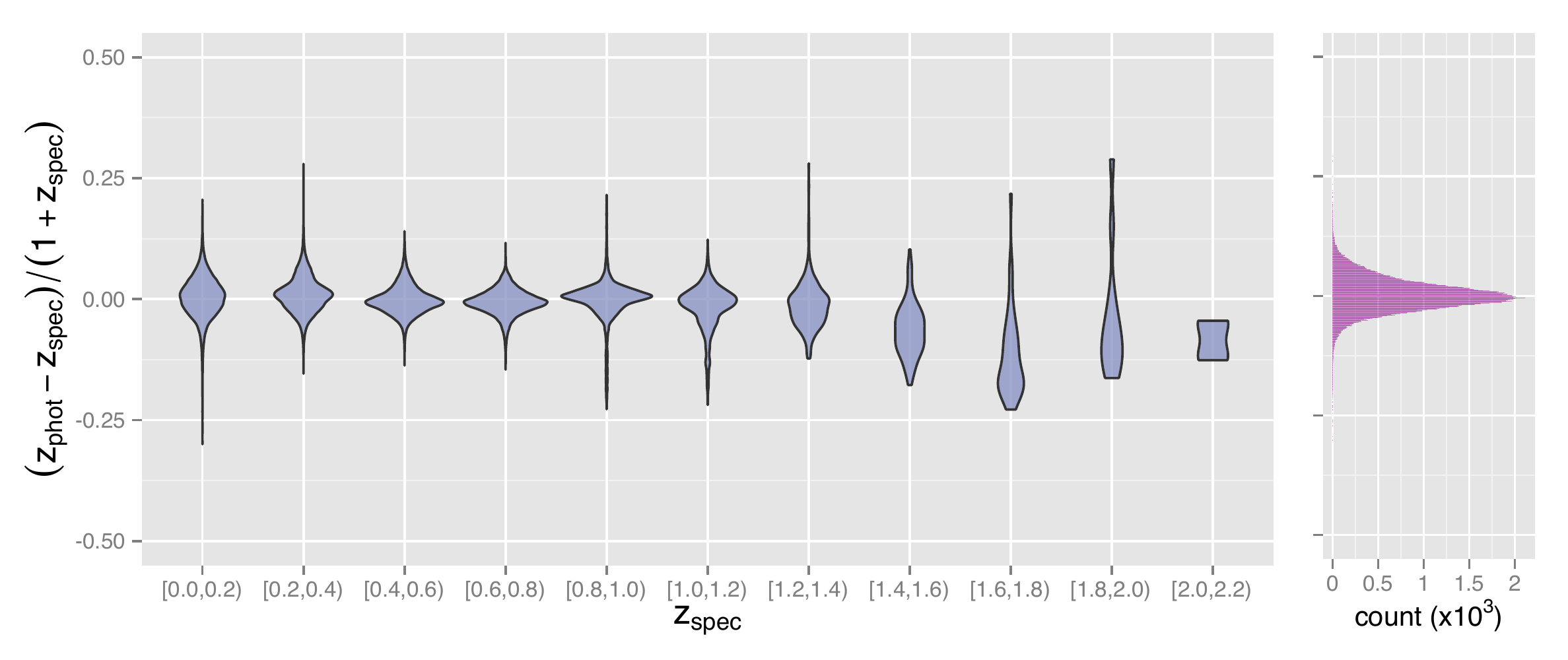}
\caption{Left panel shows the photometric redshift error distributions for the PHAT0 dataset and equation (\ref{eq:ap}) in redshift bins of width $\Delta z_{\rm spec} = 0.2$ in the range $[0 - 2.2)$. Right panel displays the error distribution all the galaxies (bins of width $\Delta((z_{\rm phot}-z_{\rm spec})/(1+z_{\rm spec})) = 0.001$).}
\label{fig:ap_figureErrDist}
\end{minipage}
\end{figure*}

\appendix
\section{Comparison with other methods} 
To compare symbolic regression with other methods and better situate our results, we adopted a publicly available data set that was previously submitted to different photo-z codes. The \textit{PHoto-z Accuracy Testing} (PHAT) was an international initiative to identify the most promising photo-z methods and guide future improvements. Two observational photometric catalogues were provided: PHAT0 with simulations, and PHAT1 with real observations. A total of 17 photo-z codes where submitted. As a direct comparison using PHAT1 is not possible, as the answers of the challenge are not openly available, we applied symbolic regression to PHAT0 and compare its results to those reported by \citet{hildebrandt2010}.

We start by splitting the original dataset, comprised by 169520 simulated galaxies in two parts: one to derive the analytical photo-$z$ expression, while another to assess the bias, scatter and outliers. For the former, in each redshift bin of $\Delta z = 0.1$ with more than 6000 objects, 3000 galaxies were randomly selected. In redshift bins with less than 6000 objects (e.g. higher redshift bins), half of the available galaxies were taken. The final subset comprises 29839 galaxies.  The remaining ones were used to assess the expression estimates. As for the SDSS-DR10 sample, we considered only the basic mathematical blocks ($+, -, *, /$), resulting in:

\begin{eqnarray}
z_{phot} &=& 0.3375 + 0.3497(r-z) + 0.3924(u-g)(Y-K) \nonumber\\
&&- (Y-J)(Y-K)- 0.4465(u-g)+ \nonumber\\
&&\frac{0.61803(J-K) + 3.4495(Y-K)(Y-J)^2}{(u-i)}.\nonumber \\
&& 
\label{eq:ap}
\end{eqnarray}

This expression, when applied to the validation data set yields $ \langle (z_{\rm phot} - z_{\rm spec})/(1+z_{\rm spec})= 0.001$,   $\sigma_{(z_{\rm phot} - z_{\rm spec})/(1+z_{\rm spec})} = 0.039$ and an outlier fraction of 4.331\%.  Here we report the outlier fraction as $\lvert z_{\rm phot} - z_{\rm spec} \rvert > 0.15 (1+z_{\rm spec})$, according to the definition adopted by \citet{hildebrandt2010}.  Results for all 17 photo-z codes submitted to PHAT for the PHAT0 data set can be summarized  as: $-0.05 \leq (z_{\rm phot} - z_{\rm spec})/(1+z_{\rm spec}) \leq 0.001$,  $0.010 \leq \sigma_{(z_{\rm phot} - z_{\rm spec})/(1+z_{\rm spec})} \leq 0.049$ and outlier fraction between 0.010\% and 18.202\%.  Comparing these results, we confirm that the accuracy of our results are within the values reported  by other widely used methods.

Finally, Figure \ref{fig:ap_figureErrDist} shows the error distributions per redshift bin. Most of the data used to derive the expression ($\approx 99.5\%$) is concentrated at $z\leq 1.45$, which not surprisingly corresponds to the interval were the photo-$z$ determination is more accurate. On the other hand, the expression shows a degraded performance at higher redshifts (which contain less than $\approx0.5\%$ of the data). This is similar to the results found for the SDSS-DR10 sample, indicating that in cases where a homogeneous data distribution is available, the symbolic regression results are competitive to available methods.


\footnotesize{

}
\end{document}